# Impact of Calcium on Transport Property of Graphene

Jyoti Katoch and Masa Ishigami*

Department of Physics and Nanoscience Technology Center, University of Central Florida, Orlando, FL, 32816

*Corresponding author:    ishigami@ucf.edu
Tel: 407-823-1155, Fax: 407-823-5112
Department of Physics, University of Central Florida
4000 Central Florida Boulevard, PS 430
Orlando, FL 32816-2385

**Abstract:**

We have measured the impact of calcium adsorbates on the transport property of graphene. Although calcium renders conductivity linearly dependent on the carrier density of graphene as predicted, our experimental results diverge from the existing theoretical calculations. Our data expose the inadequacy of any existing theory to describe the minimum conductivity of graphene and indicate that a more complete testing of the impurity scattering calculations will require improving the experimental capabilities by minimizing the contribution from the substrate-bound charged impurities and developing an ability to count the number of adsorbates while measuring transport.

Keywords: A. Graphene; D. Transport Properties; E. Adsorbate-dependent transport measurement

Graphene, a single layer of graphite, possesses unusual electronic properties characterized by relativistic Dirac physics and extraordinary field effect mobility. As such, graphene is highly useful for both fundamental science and applications [1-3]. Various extrinsic scatterers can sensitively influence the utility of graphene by obscuring the intrinsic property and affecting the performance of graphene-based electronics. As such, understanding the impact of extrinsic scatterers is essential for graphene science and technology. The impact of long-range Coulomb scatterers is most well investigated among various types of scatterers. There are two previous theory results, which define the current understanding on Coulomb scatterers. Gate-dependent conductivity for various areal densities of Coulomb scatterers has been calculated using the Boltzmann kinetic theory with random phase approximation (RPA) [4]. In addition, electron-hole asymmetry in scattering strength, unique to the Dirac physics, has been predicted for charged scatterers [5]. The previous measurements of the transport property as a function of the density of potassium adsorbates [6] showed apparently striking agreement to the theory. Potassium rendered conductivity linearly dependent on the carrier density and induced electron-hole asymmetry. Yet, sufficient discrepancies existed between the theoretical and experimental results. Quantitative differences were seen in the impact of doping by potassium and in the electron-hole asymmetry, while the behavior of the minimum conductivity was completely unexplained by the theory. These discrepancies can signal the failure in the existing theory, or they can be due to properties specific to potassium adsorbates such as density-dependent charge transfer and spatial ordering.

We have measured the impact of calcium adsorbates on the transport property of graphene. Our results parallel the previous results on potassium adsorbates. Quantitative disagreements with the existing theory are seen for the effect of doping and the electron-hole asymmetry, while the theory fails completely to describe the minimum conductivity. Minor discrepancies make it impossible to confirm or deny the validity of the theoretical calculations although they suggest that the theory at least underestimates static screening by graphene. Failure to explain the minimum conductivity is most likely due to the inability of the Boltzmann theory to describe the transport property when the Fermi wavelength is exceedingly large. Our results indicate that new experimental capabilities to minimize the contribution from the disordered substrate and to measure the number of impurities simultaneously with conductivity measurements are essential for confirming the existing theory on the impact of charged impurities on graphene.

The graphene device was fabricated from mechanically exfoliated graphene on 280 nm thermal silicon oxide on highly doped silicon using electron beam lithography. The graphene sheet was etched into the hall bar geometry using oxygen plasma after contacts were metallized [7]. The device was annealed in Ar/H$_2$ [8] to remove the resist residues prior to transport measurements performed in a ultra high vacuum (UHV) chamber, enabling a direct interaction between calcium adsorbates and the graphene sheet. Transport measurements were performed at 20 K in UHV at increasing coverage of calcium, which was evaporated from a homemade evaporator using calcium granules as the source material.

Fig. 1a shows a representative gate dependent conductivity of the graphene device at 20 K with definitions of the parameters discussed in this paper. The minimum voltage, $V_{min}$, is defined as the gate voltage at which the minimum conductivity, $\sigma_{min}$, is observed. We call this location the minimum point. Red dotted lines describe the gate dependence of conductivity away from $V_{min}$. These lines intersect above zero conductivity at the residual conductivity, $\sigma_{res}$. The plateau width is the gate voltage range where conductivity deviates from the linear behavior and the width is determined by intersecting a line through the minimum conductivity and the dotted lines. Gate-dependent conductivity, $\sigma(V_g)$, is well-described as $(1/ne\mu + 1/\sigma_c)^{-1}$ in the limit of high n, where n is the number of carriers, $e$ is the charge of an electron, $\mu$ is field effect mobility and $\sigma_c$ is a constant conductivity as observed previously [6, 9, 10]. The first term has been attributed to the scattering induced by the long-range Coulomb impurities with mobility inversely proportional to the number of impurities. $\sigma_c$ has been suspected to be due to short-range impurities or white noise impurities [9]. Field-effect mobility is assumed to be gate-independent in this paper and determined by fitting a linear line to the dependence of $n/\sigma(V_g)$ and finding its intercept at n = 0.

Fig. 1b shows measured conductivity at increasing coverage of calcium. The impact of calcium on graphene is qualitatively similar to that of potassium. Upon increasing the coverage, we observe that: $V_{min}$ shifts to more negative values, mobility is significantly reduced, conductivity is rendered linearly dependent on gate voltage, minimum conductivity varies non-monotonically, and plateau width is increased. In the following discussion, each feature will be quantitatively compared to the theoretical calculations.

The dependence of electron and hole mobility on the observed shift of $V_{min}$, $V_{shift}$, is as shown in Fig. 2. Constant evaporation rate was achieved by maintaining the same power on the calcium evaporator. In this case, the accumulated exposure time is proportional to the number of adsorbates assuming a constant sticking coefficient. The figure shows that there is a linear relationship between the number of adsorbate on the surface and inverse mobility as observed previously [6] and in consistency with the Boltzmann-RPA calculation [4].

Fig. 3a shows the dependence of $V_{shift}$ on inverse electron and hole mobility. The observed behavior is well described by a power law. The exponents are found to be 1.51 for holes and 1.37 for electrons. The Boltzmann-RPA theory calculation finds the exponents to be 1.2~1.3 [4] and finds the power law behavior to be the hallmark of nontrivial, weak screening by graphene. Yet, this agreement of the experimental and theoretical exponents is only partially relevant. The theory considered only one type of charged impurities and $V_{min}$ was assumed to be equal to $V_{shift}$. A direct comparison of the theoretical curve to our results shows an offset as shown in Fig. 3b, which is likely due to the extra charge transfer from the substrate-bound charged impurities, not considered by the theory. Shifting the theoretical curves, we find that the results are close to the theoretical curve generated using an unrealistic adsorbate-graphene distance of 1 nm as shown in Fig. 3c. The distance is approximately 3 Å for both potassium and calcium as measured and calculated previously [11, 12]. Such underestimation of $V_{shift}$ at a given scatterer concentration indicates that graphene screens calcium more effectively than expected from the Boltzmann-RPA theory calculation. Increased screening is most likely due to extra carrier density induced by the substrate-bound impurities and removal of the substrate impurities should allow more complete testing of the theory.

Electron-hole asymmetry in mobility is expected to depend on the static dielectric constants of graphene and the substrate as well as the charge transferred per adsorbate [5]. $\mu_e/\mu_h$ remains to be approximately 0.9 from before dosing to increasing coverage of calcium adsorbates as shown in Fig. 4a, indicating that both the substrate-bound Coulomb impurities and calcium adsorbates induce the same asymmetry. Fig. 4b shows the comparison of the observed asymmetry to the theoretical expectation for $\mu_e/\mu_h$. The theoretical ratio is generated using $\kappa_{substrate} = 2.45$ and $\kappa_{grapheneRPA} = 2.41$ which are dielectric constants due to the substrate and the graphene lattice on $SiO_2$ substrates [13]. As shown, decreased asymmetry for calcium compared to potassium is as expected by the theory as calcium has been calculated to transfer less charge than potassium [12]. The observed smaller ratios for both calcium and potassium can be due to underestimation of either charge transfer from adsorbates or screening by graphene. Direct simultaneous determination of both the adsorbate density and transport property should enable the elucidation of the nature of the observed asymmetry.

The impact of calcium adsorbates on $\sigma_{min}$ is poorly described by the previous theoretical calculations. $\sigma_{min}$ fluctuates near 4.5 $e^2/h$ at increasing coverage as shown in Fig. 5a, while the Boltzmann-RPA theory calculation predicts a monotonic, decreasing behavior with respect to the coverage. This nearly constant behavior seen for calcium and non-monotonic behavior seen for potassium [6] indicate the failure of the theory to describe the transport property of graphene at the minimum point. In addition, the observation of near constant $\sigma_{res}$ near 2.5 $e^2/h$, which is similar to the previous results on potassium adsorbates, is also in a complete disagreement with the theory. The Boltzmann approximation is expected to be inaccurate when the Fermi wavelength approaches infinite at the Dirac point. Yet, the Boltzmann-RPA theory has been expected to be accurate at the minimum point because the inhomogeneity introduced by the substrate-bound impurities ensures the Fermi wavelength to be finite except at the few points where the Fermi level crosses the Dirac point. The observed $\sigma_{min}$ and $\sigma_{res}$ demonstrate that these crossing points are sufficient to make the Boltzmann-RPA theory inadequate in describing the minimum point.

Unlike $\sigma_{min}$ and $\sigma_{res}$, the plateau widths show qualitative agreement to the Boltzmann-RPA theory. The theoretical result remains closer to our experimental results away from the minimum point. The observed widths are smaller than the theoretical values in contradiction with the Boltzmann-RPA theory as the charged scatterers bound on the substrates should be contributing to widen the plateau. Such discrepancy can be due to the experimental inaccuracy in determining the plateau. Minimizing the contribution from the substrate impurities should enable better experiments.

In conclusion, we have measured the impact of calcium adsorbates on the transport property of graphene to determine if the discrepancies seen in the previous measurements using potassium adsorbates are systematic to all charged impurities or specific to the interaction between potassium and graphene. We find that similar behavior is manifested by graphene under the influence of calcium. Our results indicate that the Boltzmann-RPA theory has not been properly confirmed by experiment and is inadequate at the minimum point. Further tests on the previous theoretical calculations must minimize the influence of the substrate as have been achieved by hexagonal boron nitride [14] and count the density of charged scatterers while measuring their impact on the transport property.

**Acknowledgements:**

This work is based on research supported by the National Science Foundation under Grant No. 0955625.

**Figure Captions**

**Figure 1** (a) Gate dependent conductivity of the graphene device used for the experiment. The figure shows the definition for the plateau width, $V_{min}$, minimum conductivity ($\sigma_{min}$) and residual conductivity ($\sigma_{res}$). Dotted red lines are used to determine the plateau width and residual resistivity. (b) Gate dependent conductivity of the graphene device at increasing levels of calcium adsorbates.

**Figure 2** Inverse electron and hole mobility as a function of the calcium dosage time.

**Figure 3** (a) $V_{shift}$ as a function of inverse electron and hole mobility. A power law behavior is observed for both electron and hole mobility. (b) $V_{min}$ as a function of inverse electron and hole mobility. Solid lines are calculated values for charged impurities located 0.3 nm and 1 nm away from graphene. (c) Theoretical curves have been offset in the x-axis by 1.4 sec/V m$^2$.

**Figure 4** (a) $\mu_e/\mu_h$ at increasing coverage. (b) Theoretical $\mu_e/\mu_h$ at different Z for adsorbates. Green and brown dots indicate the experimental values for calcium and potassium. Values of charge transfer for potassium and calcium are as calculated previously [12].

**Figure 5** Comparison of observed data and theory of (a) minimum and residual conductivity and (b) plateau widths.

# References

1. A.K. Geim, Science, 2009. **324**: p. 1530.
2. A.K. Geim and K.S. Novoselov, Nature Materials, 2007. **6**: p. 183.
3. K.S. Novoselov, D. Jiang, F. Schedin, T.J. Booth, V.V. Khotkevich, S. Morozov, and A.K. Geim, *Two-dimensional atomic crystals.* Proc. Natl. Acad. Sci. U. S. A., 2005. **102**(30): p. 10451.
4. S. Adam, E.H. Hwang, V.M. Galitski, and S. Das Sarma, *A self-consistent theory for graphene transport.* Proceedings of the National Academy of Sciences of the United States of America, 2007. **104**(47): p. 18392-18397.
5. D.S. Novikov, *Numbers of donors and acceptors from transport measurements in graphene.* Applied Physics Letters, 2007. **91**(10): p. 102102
6. J.H. Chen, C. Jang, S. Adam, M.S. Fuhrer, E.D. Williams, and M. Ishigami, *Charged-impurity scattering in graphene.* Nature Physics, 2008. **4**(5): p. 377-381.
7. *Contacts were thermally evaporated gold with chromium sticking layers.*
8. M. Ishigami, J.H. Chen, W.G. Cullen, M.S. Fuhrer, and E.D. Williams, *Atomic structure of graphene on SiO2.* Nano Letters, 2007. **7**(6): p. 1643-1648.
9. J.H. Chen, C. Jang, M. Ishigami, S. Xiao, W.G. Cullen, E.D. Williams, and M.S. Fuhrer, *Diffusive charge transport in graphene on SiO2.* Solid State Communications, 2009. **149**(27-28): p. 1080-1086.
10. C. Jang, S. Adam, J.H. Chen, D. Williams, S. Das Sarma, and M.S. Fuhrer, *Tuning the effective fine structure constant in graphene: Opposing effects of dielectric screening on short- and long-range potential scattering.* Physical Review Letters, 2008. **101**(14): p. 146805.
11. M. Caragiu and S. Finberg, *Alkali metal adsorption on graphite: a review.* Journal of Physics: Condensed Matter 2005 **17**: p. R995.
12. K.T. Chan, J.B. Neaton, and M.L. Cohen, *First-principles study of metal adatom adsorption on graphene.* Physical Review B, 2008. **77**: p. 235430.
13. E.H. Hwang and S. Das Sarma, *Dielectric function, screening, and plasmons in two-dimensional graphene.* Physical Review B, 2007. **75**: p. 205418.
14. C.R. Dean, A.F. Young, I. Meric, C. Lee, L. Wang, S. Sorgenfrei, K. Watanabe, T. Taniguchi, P. Kim, K.L. Shepard, and J. Hone, *Boron nitride substrates for high-quality graphene electronics.* Nature Nanotechnology, 2010. **5**(10): p. 722.

Figure 1

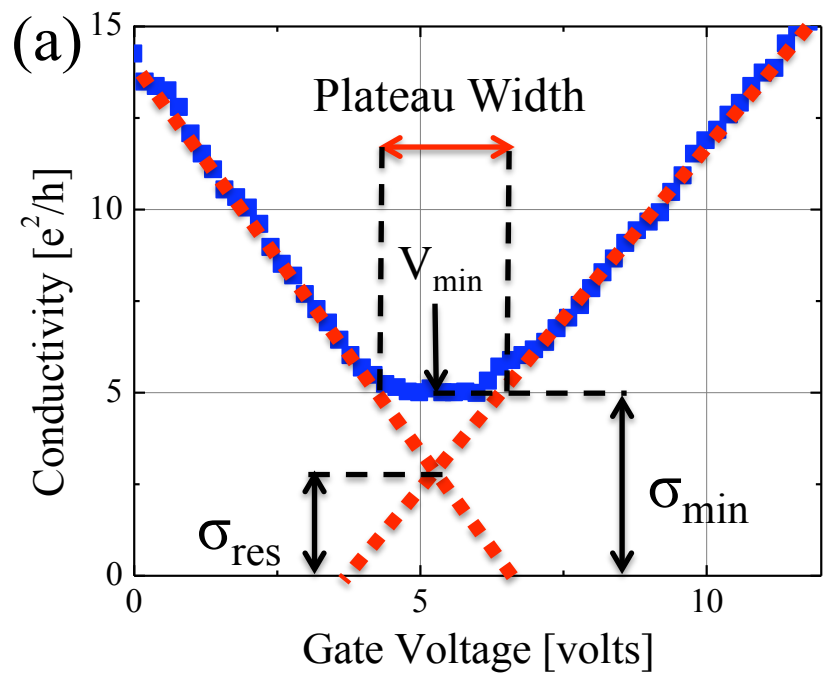 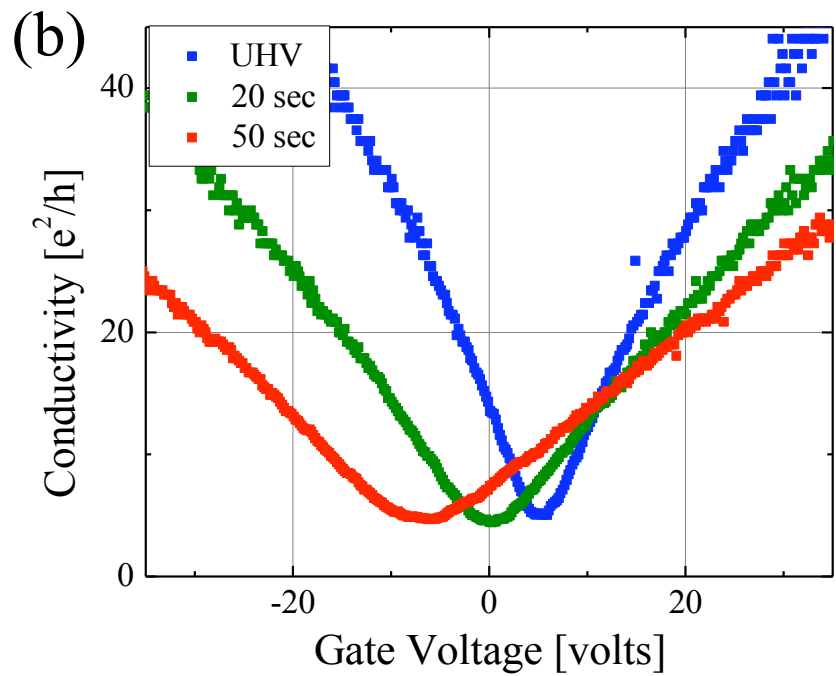

Figure 2

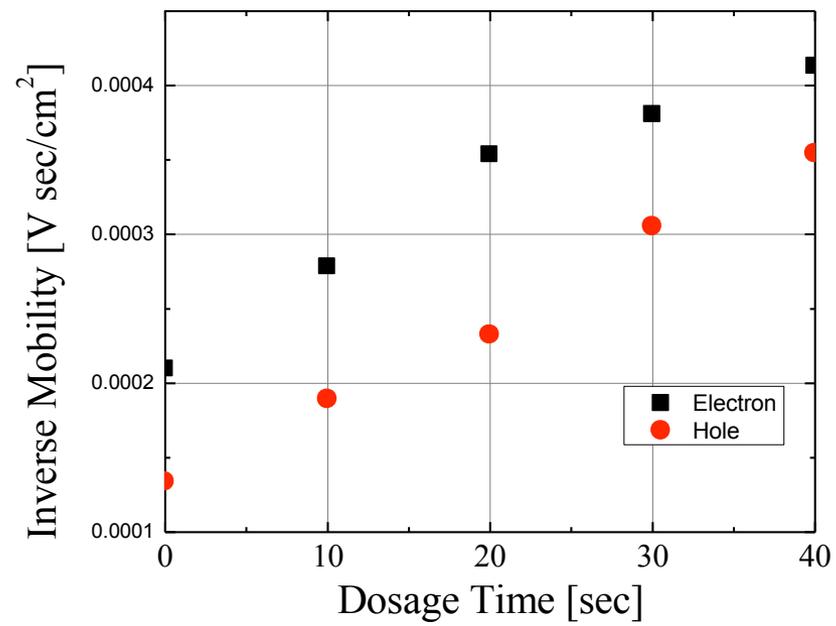

# Figure 3

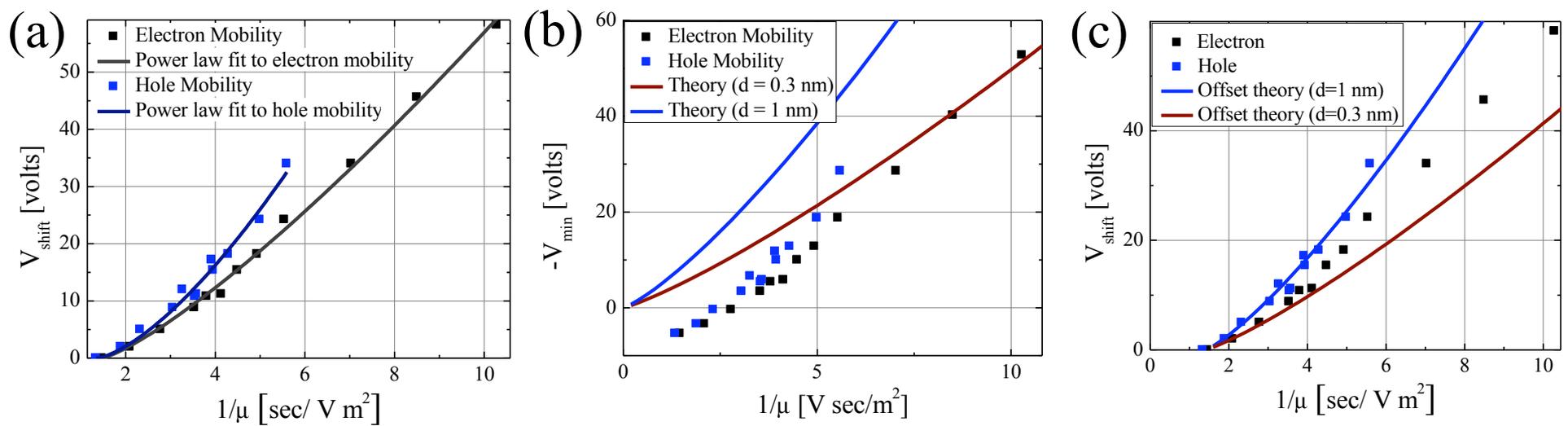

Figure 4

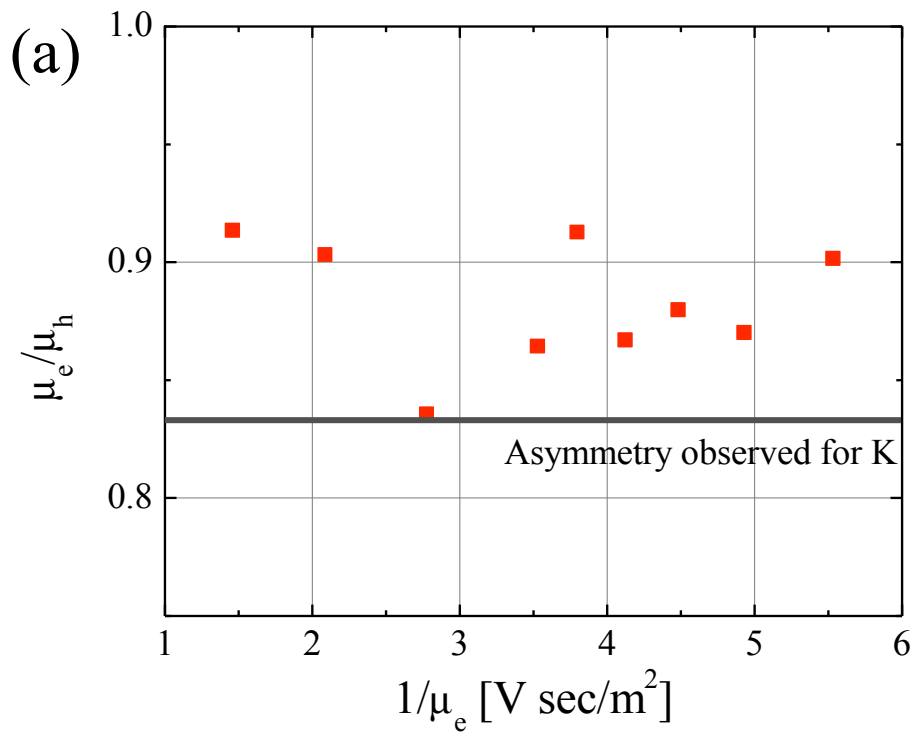 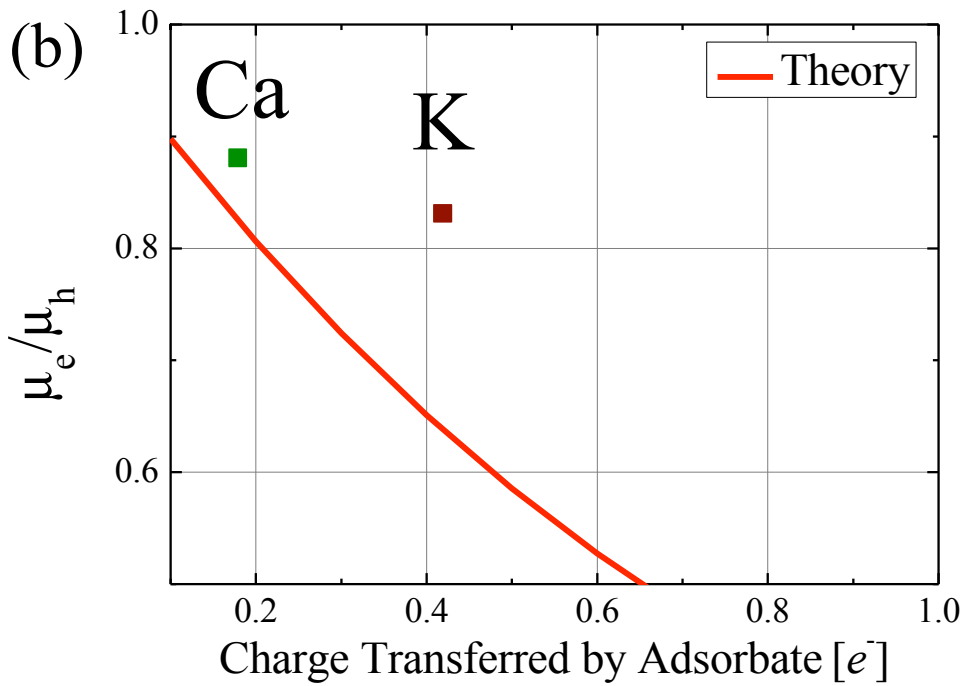

Figure 5

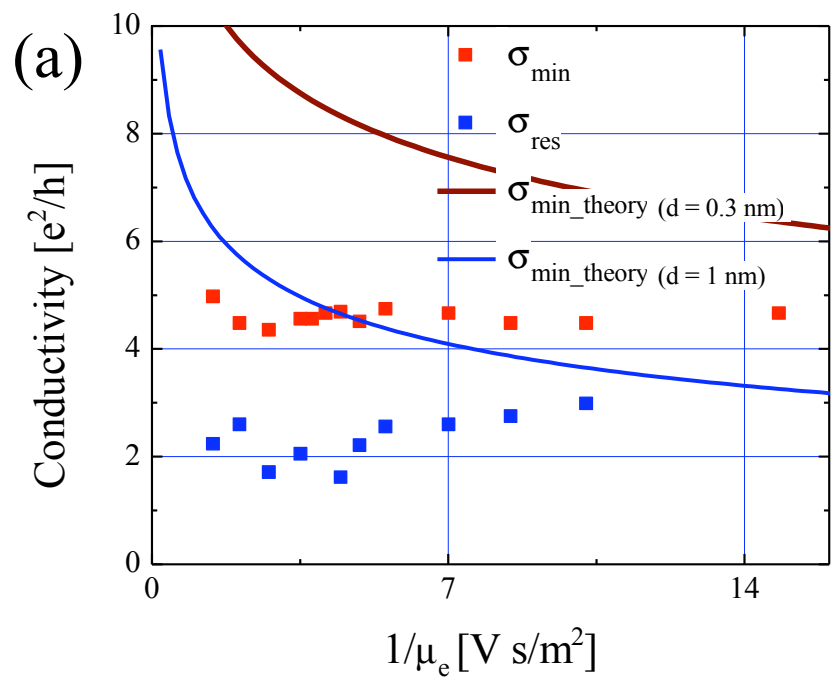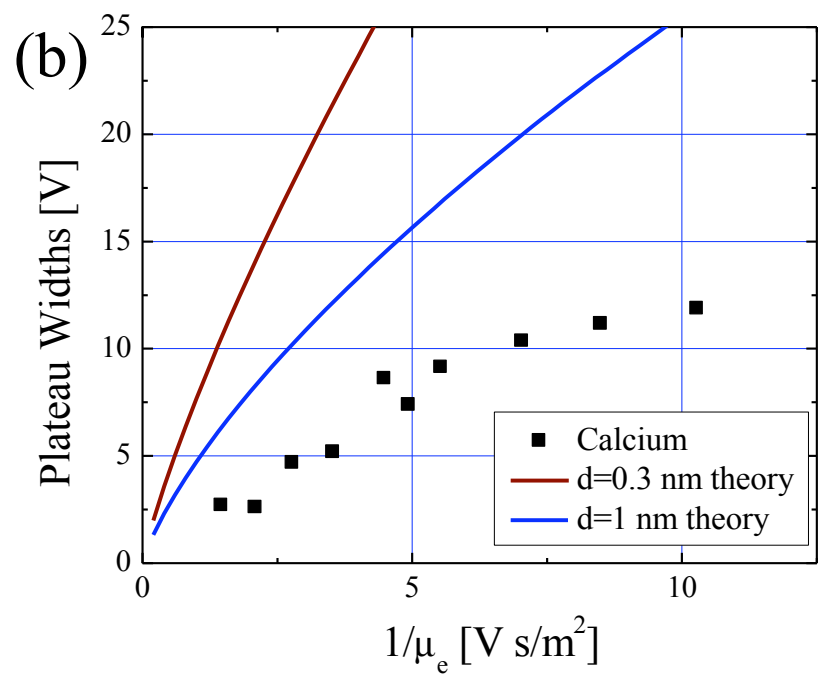